\documentclass[twocolumn,pre]{revtex4}

\usepackage{graphicx}
 \usepackage{rotating}
\usepackage{amsmath}
\usepackage{amsfonts}
\usepackage{amssymb}
\usepackage{graphicx}
\usepackage{enumerate}
\usepackage{longtable}
\usepackage{dcolumn}
\usepackage{bm}
\usepackage{epsfig}
\usepackage{color} 
\begin{document}

\title{Analysis and modeling of  low frequency local field oscillations in a hippocampus circuit under osmotic challenge: the possible role of arginine vasopressin circuit for hippocampal function.}
\author{Hern\'an Barrio Zhang $^1$}
\author{Mariana M\'arquez-Machorro$^5$}
\author{Vito S. Hern\'andez $^2$}
\author{Andr\'es Molina $^2$}
\author{Limei Zhang $^2$}
\author{Tzipe Govezensky$^3$}
 \author{Rafael A. Barrio$^{4}$}
 \email{barrio@fisica.unam.mx} 
 
 \affiliation{$^1$Facultad de Ciencias, U.N.A.M., 01000 - M\'exico D.F., Mexico.}
  \affiliation{$^2$Facultad de Medicina, Departamento de Fisiolog\'ia, Universidad Nacional Aut\'onoma de M\'exico, Mexico.}
   \affiliation{$^3$Instituto de Investigaciones Biom\'edicas, Universidad Nacional Aut\'onoma de M\'exico, 04510 M\'exico D.F., Mexico,}
\affiliation{$^4$Instituto de F\'isica, U.N.A.M., Circuito de la Investigaci\'on Cient\'ifica Ciudad Universitaria, C.P. 04510, M\'exico D.F., Mexico,}
\affiliation{$^5$McGill University, Canada.}

\date{\textrm{\today}}

\begin{abstract}
Electrophysiological time series were taken simultaneously in two locations in the hippocampus of a rat brain previously described as receiving innervation from the osmosensitive vasopressinergic neurons of the hypothalamus. A hyperosmotic saline solution injection was administered during the time of the experiment.  We analyze the recorded time series using different methods. We detect a modification of the delta and theta oscillations just after the perturbation caused by the injection. We compare the quality and information that each one of the methods exhibit and we analyze the characteristics of the perturbation based on a hypothesis that the strength of the functional connections between the vasopressinergic hypothalamic magnocellular neurons and their target in the hippocampus is modified by the perturbation. We built a model of the hypothetic neural connections and numerically calculate the time series produced by the system when simulating the perturbation caused by the saline injection. The theoretical results resemble the experimental findings concerning the frequency and amplitude alterations of the delta and theta bands.
\end{abstract}

%\pacs{71.27.+a,75.30.Ds}
\maketitle

%\date{Received: date / Revised version: date}
\section{Introduction}
\label{intro}

The study and analysis of time series of recorded data is a powerful tool to investigate the dynamics and interactions of physical, chemical or biological processes in many and diverse systems. This is particularly true in the study of the brain functions, since the brain activity can be measured by the collective electrical activity of  neurons and their activation or inhibition by their synaptic connexions and other means of communication.

This electrical activity is usually detected by encephalograms, MRI, or more precise electrophysiological recordings. The time series collected this way contain much information about particular time-space correlations of the electrical activity of the brain. It is therefore important to use the  appropriate methods of analyzing these data  in order to maximize the amount of information that could be extracted from them.

The interactions between different parts of the brain are regulated by many substances called neuromodulators that certainly modify the transfer of signals from one neuron to the other. An important neuromodulator is vasopressin, also called the antidiuretic hormone, that is known to control water homeostasis and blood pressure. It is released by the pituitary into the blood stream, but also directly to the brain, where it is believed to play an important role in social behavior, sexual conduct, motivation, learning and memory, and response to stress. 

In this work we record electrophysiological time series simultaneously from two specific parts of the brain that are known to contain vasopressinergic innervation from the hypothalamus and that could be modified by perturbing the hydroelectrolitic homeostasis. More specifically, we want to study the time varying properties of non stationary changes in the local field potentials and the coherence in the dorsal and ventral regions of the hippocampus of a rat. The signals are taken simultaneously in both regions and the rat subjected to the activation of the hypothalamic vasopressinergic system by a systemic injection of a hyperosmotic saline solution.

We use different approaches to analyze the data and compare the information extracted from each one of them. We also develop a theoretical model based on the assumption that the external perturbation modifies the communication pathways between the regulating system localized in the hypothalamus and the  hippocampal regions.

\section{Biological facts}
\label{intro}
Central nervous system neurons exchange information via electrical currents, the sum of all the electrical currents generates changes in the electrical potential of the extracellular medium that vary in time. The low frequency ($< 500 $ Hz) component of these signals is known as the local field potential (LFP) and can be recorded and studied from a local network of neurons. Fluctuations in the amplitude of the LFP can be measured in different parts of the brain~\cite{Caton}, and recent findings indicate that network oscillations bias input selection, temporally link neurons into assemblies, and facilitate synaptic plasticity, mechanisms that cooperatively support temporal representation and long-term consolidation of information~\cite{Buzsaki}.

	Using power spectra, LFP oscillations have been divided into several arbitrarily defined frequency bands, including the internationally agreed delta (0.1 - 4 Hz), theta (4-8 Hz), alpha (8-15 Hz), beta (15-30 Hz) and gamma (30 -80 Hz) bands~\cite{Noachtar}. The presence and power of these frequency bands have been associated with different mental states, tasks and behaviors~\cite{Buzsaki2}. However, there are some limitations in this classification since some physiologically relevant rhythms sometimes fall into two categories, for example in the awake rodent, hippocampal theta oscillations fall between 4 and 10 Hz~\cite{Buzsaki2006}. 

	The hippocampus is one of the most studied regions of the brain, one of its characteristics is the presence of prominent theta oscillations related to: REM sleep, spatial representations, attention, arousal or anxiety~\cite{Lever}. The hippocampus can be divided in functionally different, albeit connected, dorsal (dHi) and a ventral (vHi) subregions mediating learning/memory and emotional control/stress responsivity functions, respectively~\cite{Fanselow}. Some recent studies suggest that theta oscillation coherence between these subregions may increase during stressful situations~\cite{Gray}.  

\begin{figure}[ht!]
\begin{center}
\includegraphics[width=\columnwidth]{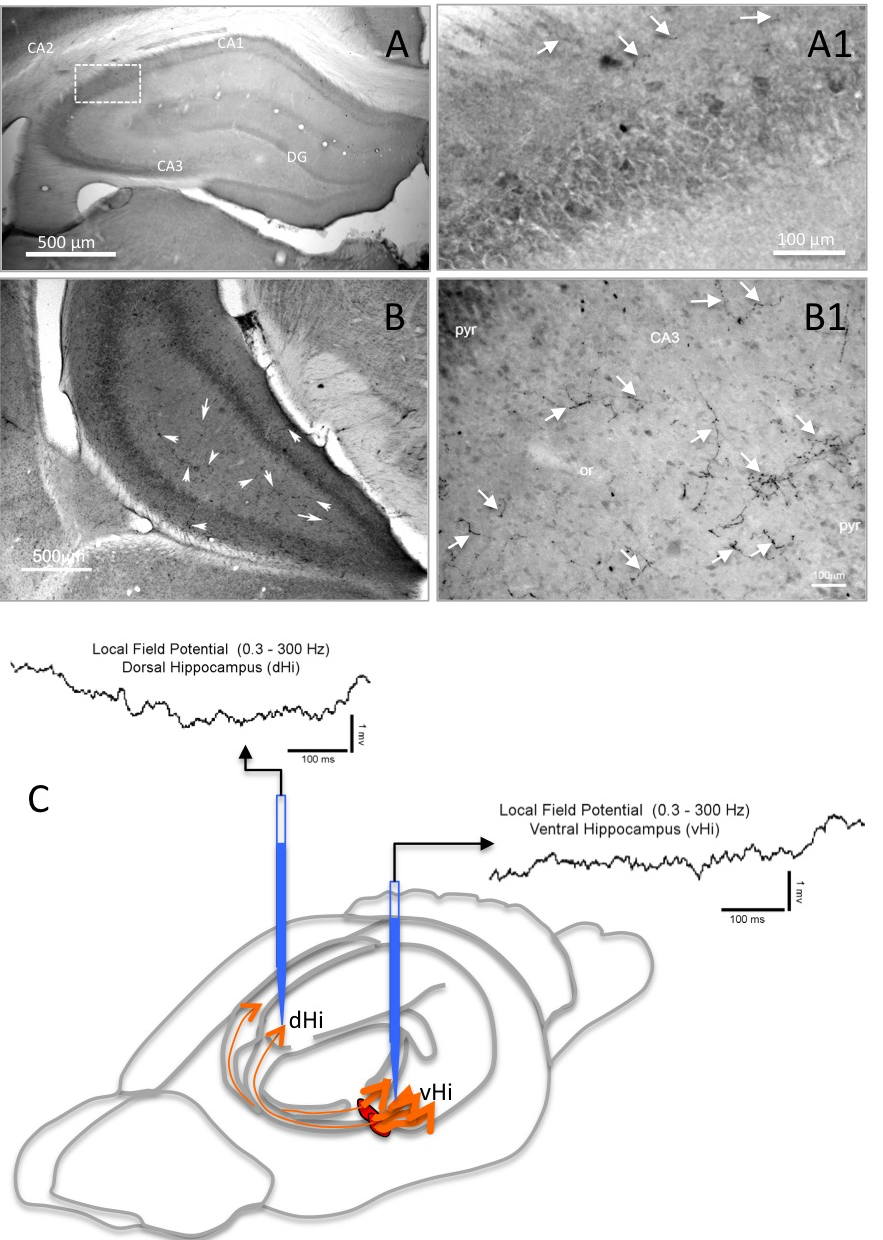} %fig1new.pdf
\caption{The dorsal (dHi) and ventral (vHi) hippocampus are innervated by vasopressinergic fibers originated in the hypothalamic paraventricular (PVN) nuclei. A and A1 Low and high power microphotograph of dorsal hippocampus showing fibers (arrows) immunolabelled with an antibody against vasopressin. B and B1: Low and high power microphotograph of ventral hippocampus showing fibers (arrows) immunolabelled as mentioned. C:  Schematic representation of the vasopressinergic pathways (orange arrows) by which vasopressin fibers originating in the magnocellular neurons of the PVN (Red ovals) reach the dorsal (dHi) and ventral (vHi) hippocampi, the arrow thickness represent the density  of fibers in each of these pathways. The sites of LFP recordings in the dorsal and ventral hippocampus are indicated.}
\label{fig0}
\end{center}
\end{figure}

	We have previously demonstrated that magnocellular neurons located in the supraoptic and paraventricular nuclei of the hypothalamus , known by their key role in the homeostatic control of hydroelectrolite balance, in addition to their peripheral axonal projections, send ascending projections to intracerebral limbic targets including the dorsal (dHi) and ventral (vHi) hippocampus~\cite{Zhang,H1,H2}  (see Fig.~\ref{fig0}),  this finding suggest that the hypothalamus could modulate the activity of local neuronal networks in the dHi and in the vHI, modifying the oscillatory activity and the coupling between the dHi and vHI and thus have an influence in the integration of the cognitive-emotional functions of these structures. 
	
\section{Methods}
\subsection{Animals}

Experiments were performed on adult male Wistar Rats (280-300g), provided by the local animal facility and housed at $20-24^{\circ}$C on a 12h dark/light cycle (lights on at 19:00h) with tap water and standard rat chow pellets available ad libitum. All surgical and experimental procedures were performed in accordance with the guidelines published in the National Institutes of Health Guide for the Care and Use of Laboratory Animals (publication number 86-23, revised 1987) and with the approval of the local bioethical and research committee (CIEFM-086-2013).

\subsection{Electrophysiological recordings} 

For in vivo extracellular recording, rats were induced into anesthesia with 4\% isourane in oxygen, followed by urethane injection (intraperitoneal, 1.3 g/kg, Sigma-Aldrich), with supplemental doses of xylazine (30 mg/kg), as necessary. Body temperature was maintained at $36^{\circ}$C with a heating pad. 

	Once anesthetized, animals were placed on a stereotaxic frame and  two craniotomies  were performed in order to position the tip of two glass electrodes (8-15 MOhms) coated with DiI (a lipophilic fluorescent dye that allowed to confirm the correct positioning of the electrodes) at previously standardized coordinates\cite{Paxinos} in dHi and vHi, selected based on our previous work defining the main sites of vasopressinergic innervation from the hypothalamic vasopressinergic neurons: CA1/CA2 region of dorsal hippocampus (coordinates : -2.3 mm posterior to Bregma, 2 mm lateral to midline and 3 mm ventral to skull surface) and CA1 ventral hippocampus region  (ipsilateral coordinates: -4.8 mm posterior to Bregma, 5 mm lateral to midline and 8.5 mm ventral to skull surface). Glass monopolar recording electrodes were referenced against a wire implanted subcutaneously in the neck. Extracellular LFP signals from both electrodes were amplified (ELC-01MX amplifier, npi electronics, GmbH, Tamm, Germany), filtered between 0.3 and 500 Hz (BF-48DGX  Filter, npi electronics), digitized at 1000 Hz (INT-20X breakout box module connected with a National Instruments, NI-M series board) and displayed with the Sciworks Experimenter software (Datawave technologies)

\subsection{Osmotic Stimulation}

To activate the hypothalamic vasopressinergic system, after at least 20 minutes of stable basal LFP recording had elapsed, a 2\% body weight volume of a 900 mM NaCl solution was intraperitoneally injected. Data recording continued during at least 45 minutes after the NaCl injection.

We performed numerous experiments following the above mentioned conditions. For illustration purposes we chose a representative set of data.

\subsection{Histological examination}

After the experiment, the rats were given a lethal dose of sodium pentobarbital (63 mg/kg, Sedalpharma, Mexico), and transcardially perfused with isotonic NaCl followed by paraformaldehyde fixative (4\% paraformaldehyde, 15\% v/v picric acid in 0.1M phosphate buffer), the brains were collected, coronally sectioned at 70 $\mu$m in a vibratome (Leica VT-1000s). Sections were mounted in microscope slides and were observed under fluorescence microscopy to verify the correct positioning of the electrodes in dHi and vHi. 

\section{Time series analysis}

In this section we explain the methods used to analyze the experimental data. For the sake of clarity, we  show results extracted from a single set of representative Data.  These were taken every millisecond  thus each record consists of 3 million number pairs. 
 Fig.\ref{fig0}E is a representative diagram to show the position of the electrodes recording  data of series v1$(t)$ (dHi) and v2$(t)$ (vHi).

In order to see the distribution of frequencies in the data the first thing to do is to calculate the Fourier power spectra of the time series, which is defined as,

\begin{equation}
\label{fp}
\mathrm{Power\;v_i} (\omega)=\left | \int_{t_i}^{t_f} \mathrm{v_i}(t)e^{iwt}dt \right |
\end{equation}
where the integral is taken from the initial time $t_i$ to the final time $t_f$.

\begin{figure}[ht!]
\begin{center} 
\includegraphics[width=\columnwidth]{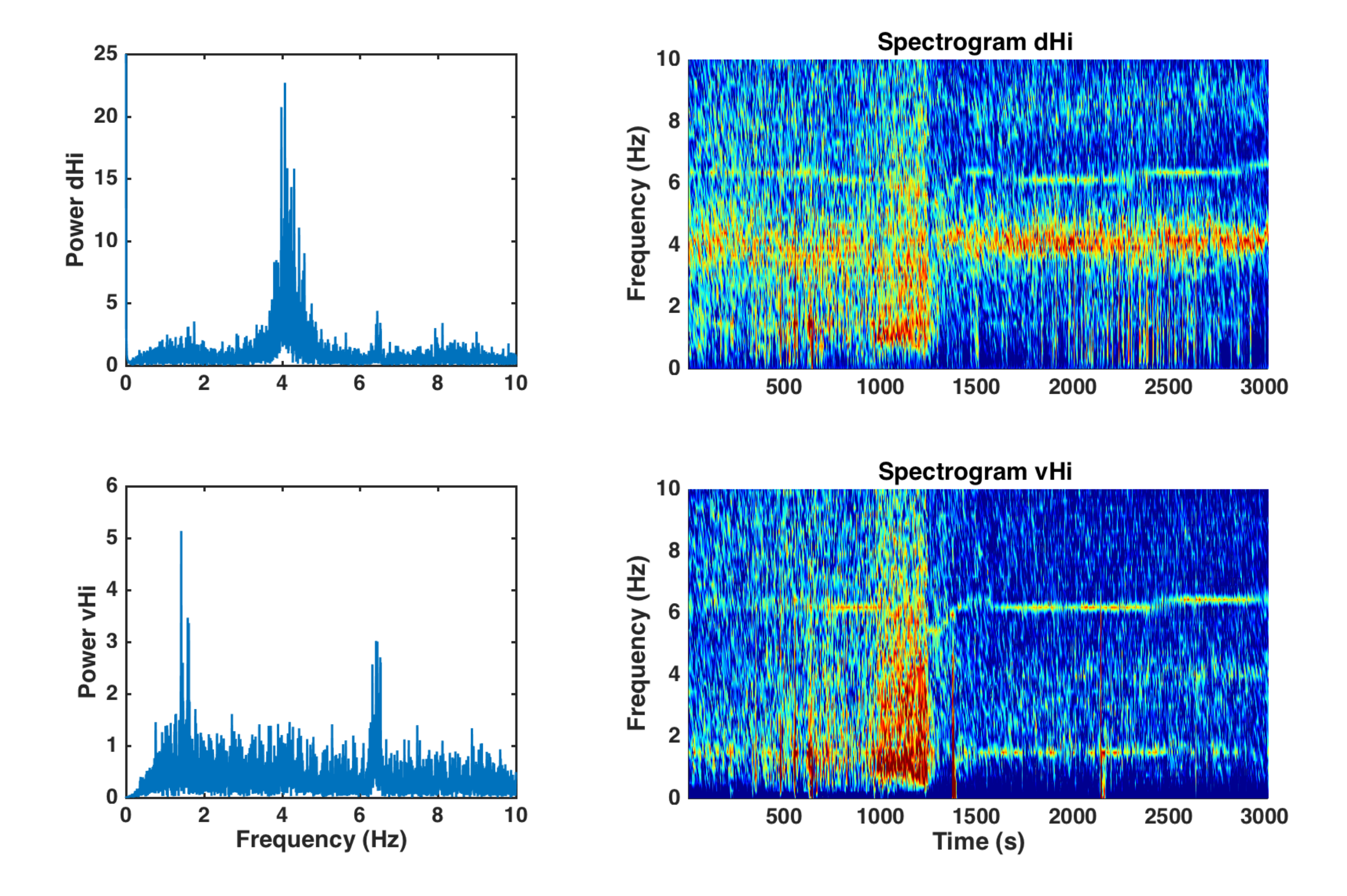}
\caption{Fourier power spectra and corresponding spectrograms of the time series v1 and v2.}
\label{fig:t2}
\end{center}
\end{figure}

In the left column of Fig.~\ref{fig:t2} we show the power spectra of the filtered data. Notice that in Power dHi there is a broad bands with a conspicuous peak at $\sim 4$ Hz,while in the vHi spectrum there are  two peaks at around 2 and 6 Hz. However,  these spectra  are calculated supposing the series are stationary and do not contain information about the temporal appearance and coherence of these bands in the system. 

In order to preserve time information one can use the windowed Fourier transform to construct the so called spectrogram. This is basically a three dimensional plot formed with a succession of $n$ Fourier power spectra taken from a small time window $w(j)=[ t_i(j), t_f(j)]$. 

Then, the frequency dependent power spectra are plotted in the $y-z$ plane, and the successive $n$ spectra are stacked along the $x$ axis, which then contains the time windows as they are sliding along time (notice that the windows could overlap or not). The three dimensional plot is usually showing the $x-y$ plane and the $z$ plane is represented with a color code. In the right column of Fig.~\ref{fig:t2} we show the spectrograms of the two time series using the Fourier spectra of the data. In all spectrograms shown here we used a window of $w(j)=4096$ s, with an overlap of 2048 s.

\begin{figure}[ht!]
\begin{center} 
\includegraphics[width=\columnwidth]{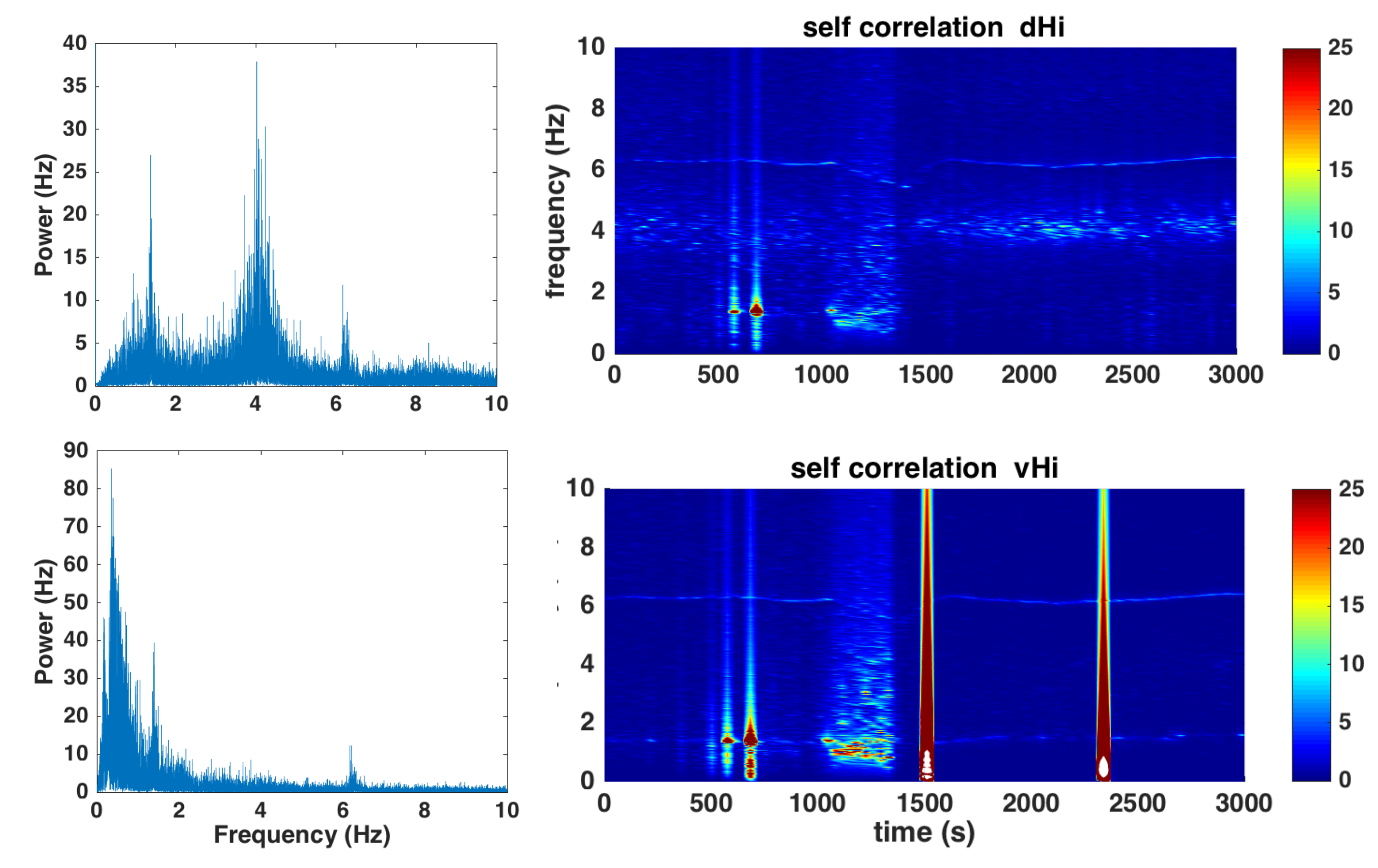}
\caption{Power spectra and spectrograms of the self correlations of the same data as in Fig.~\ref{fig:t2}.}
\label{fig:t3}
\end{center}
\end{figure}

A better way to investigate repeating patterns in the data is to calculate the self correlation of the series, defined as
\begin{equation}
\label{R}
R_{ff}(\tau)=\int_{-\infty}^{\infty} f(t)f^*(t-\tau)dt=\int_{-\infty}^{\infty} f(t+\tau)f^*(t)dt.
\end{equation}
 and then calculate the power spectra of the self correlation. The self correlation emphasizes the real oscillations over the background. In the left column of Fig.~\ref{fig:t3} we show the calculation of the self correlation power spectra for the two series. Notice that now the frequency bands are better defined, with a good separation between bands, now the bands at 2 Hz and 6Hz (only hinted in the Fourier spectrum of dHi) are clearly visible, and a band at very low frequency appears in vHi.

In the right column of Fig.~\ref{fig:t3} we notice that the spectrogram obtained with the self correlation spectra is much clearer than than the one obtained with the original data.  In both we notice that the particular external event at $t_p=1000$ seconds (the injection) caused a substantial change in the bands. 

The low frequency band ($ \sim 1$ Hz at times 500, 700 1500 and 2400 s), is due to external noise during the experiment. This is clearly noticeable in the spectrogram of vHi. In dHi the band at 4 Hz disappears with the perturbation and the band at 2 Hz (not normally present) appears after the perturbation. The band at  $ \sim 6$ Hz (theta oscillations), which is evident in both series, is perturbed by the injection lowering its frequency to 5 Hz and then recovering  with a rebounce to higher frequencies, after some 400 s.  This is clearly seen in both series and in the Fourier and self correlation spectrograms. The effect of the perturbation is only seen between $t_p=1000$ s and $t_r=1300$ s. The interval from $t_p$ to $t_r$  should be considered as the time the system takes to reach a new stable state after the injection.
 
 The cross correlation of the two time series will detect correlated signals, it detects periods of simultaneous behavior, or when there is a delay between the same behavior in both signals. This quantity is defined as 
  \begin{equation}
\label{C}
C_{\mathrm{v1,v2}}= \int_{-\infty}^{\infty} \mathrm{v1}(t)\mathrm{v2}^*(t+\tau)dt.
\end{equation} 

\begin{figure}[ht!]
\begin{center} 
\includegraphics[width=\columnwidth]{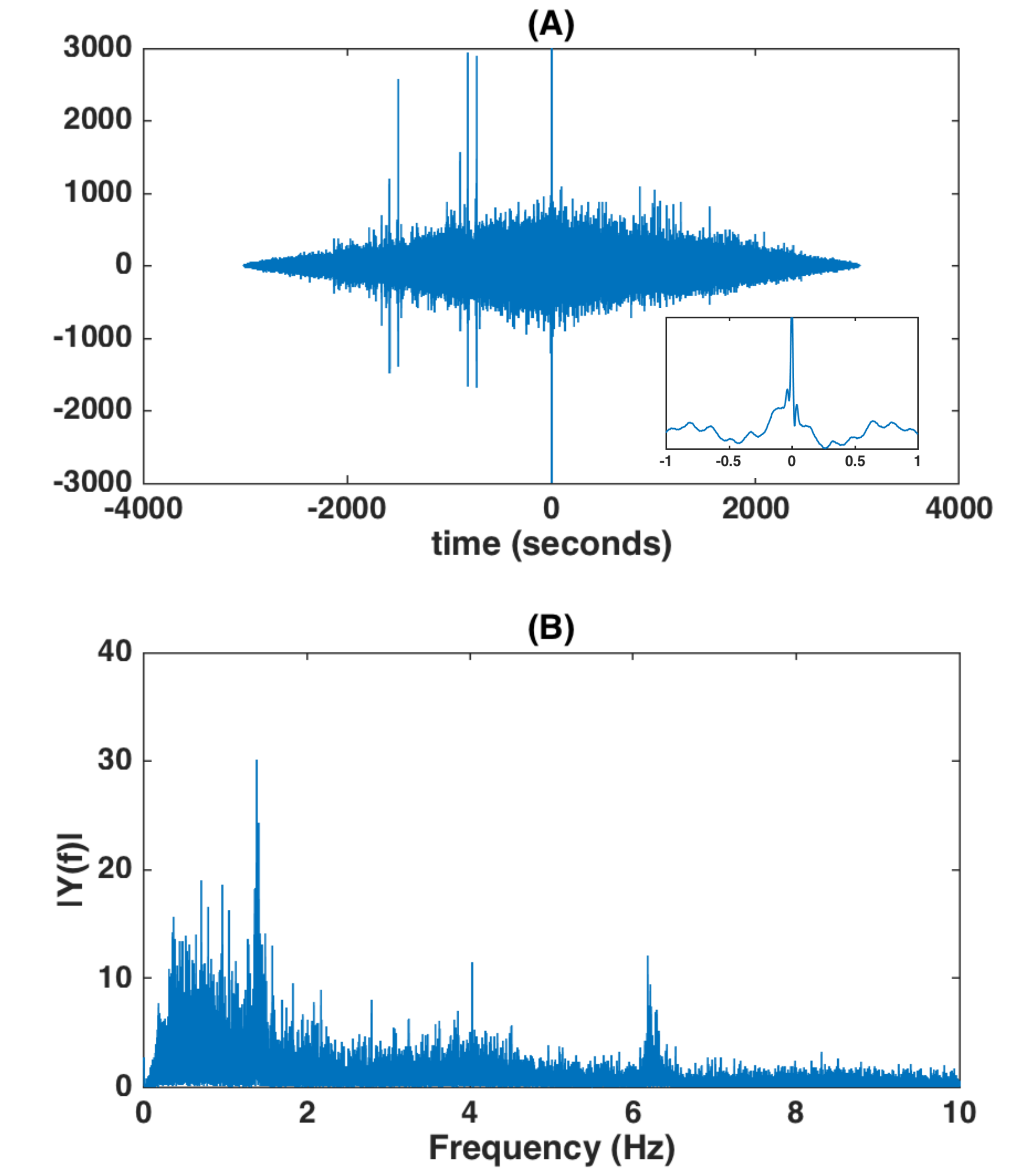}
\caption{(A) Cross correlation function of the filtered data. In the inset we show a small interval of time around zero, to illustrate the asymmetry of the cross correlation. (B) The corresponding power spectrum. }
\label{fig:t5}
\end{center}
\end{figure}
 In the upper panel of Fig.~\ref{fig:t5} we show the cross correlation function $C_{\mathrm{dHi,vHi}}$. Notice that the plot is almost symmetric for positive and negative time differences. The greatest correlation is apparently at zero, indicating that changes were almost simultaneous. However, in the inset we show and amplification around zero to verify that there is a small delay of the cross correlation of the order of some milliseconds. In the lower panel we show the power spectrum of the cross correlation function. The three low frequency bands are clearly noticeable. 
 
\begin{figure}[ht!]
\begin{center} 
\includegraphics[width=\columnwidth]{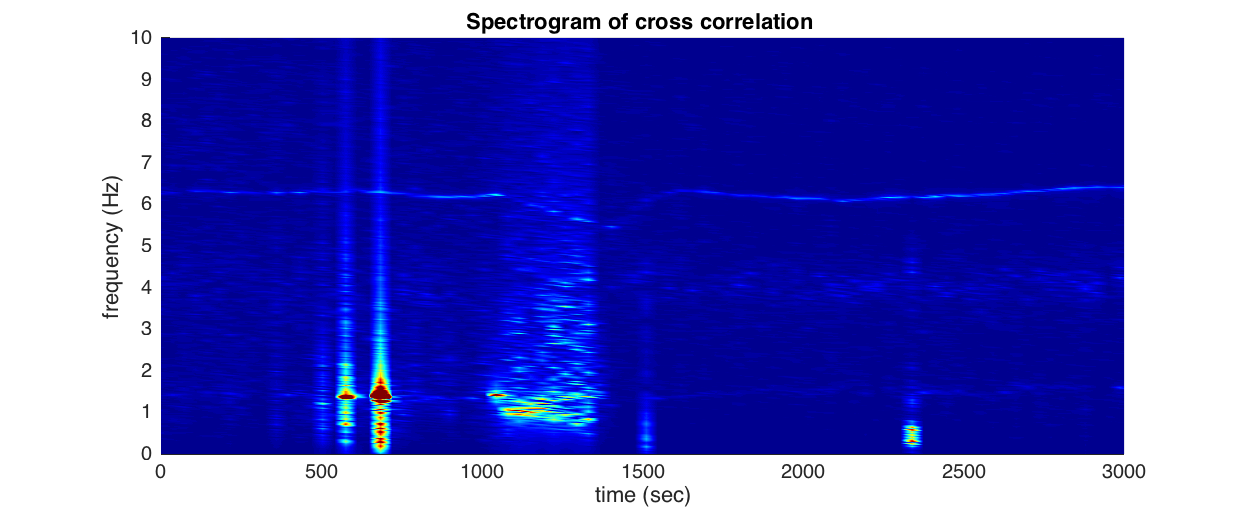}
\caption{Spectrogram using the cross correlation function of the filtered data. The color scale is in this case linear.}
\label{fig:t6}
\end{center}
\end{figure}

Our conclusion is that the power spectrum of the cross correlation is a better tool to investigate the main recurrent frequencies present in  both series, since neurophysiological signals usually repeat in different places with some time delays, which seems to be the case here. In order to acquire time information, as well as synchronous or delayed frequencies, one can construct the spectrogram with the cross correlations instead of the self correlations. In Fig.~\ref{fig:t6} we show such a calculation. Notice that the information given by the self correlation spectrograms is seen more clearly in this figure, including very small disturbances in the $ \sim 2$ (delta) and 6 Hz (theta) bands.

The final method used is to examine scalograms obtained with wavelets. The continuous wavelet transform, unlike the Fourier transform, is able to represent the data in a time-frequency domain very precisely, and it does not require the assumption of stationarity. It is defined as a usual convolution,

\begin{equation}
\label{wt}
W_{\mathrm{v}}(\phi,a,b)=\frac{1}{\sqrt{a}}\int_{-\infty}^{\infty}\mathrm{v}(t) \phi^* \left( \frac{t-b}{a}\right)dt,
\end{equation}
where $\phi$ is a continuous function of time called the mother wavelet, $a$ is a scale, and $b$ is a time translation.

In this work we have used the so called Morlet mother wavelet, whose  Fourier transform is defined as
\begin{equation}
\label{mw}
\phi(s\omega)=\frac{1}{\pi^{1/4}}e^{-(s\omega-\omega_0)^2/2}\Theta(s\omega),
\end{equation}
where $\Theta$ is the step (Heaviside) function, $s$ is a scale that should be larger than twice the sampling period, and $\omega_0$ is the mean frequency. In our case we chose zero mean. An important issue in wavelet analysis is the correct choice of scales, which depends on the frequency range one wants to be focusing and on the time resolution one desires.

\begin{figure}[ht!]
\begin{center} 
\includegraphics[width=\columnwidth]{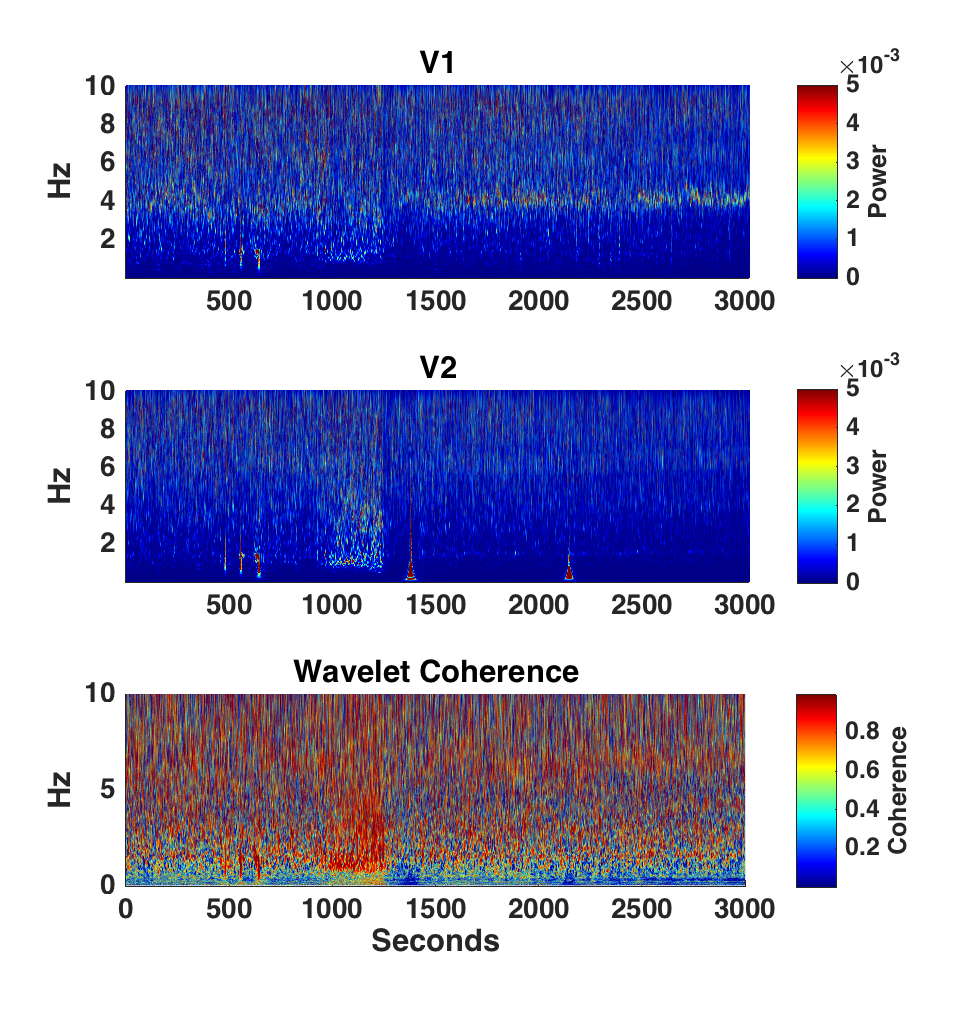}
\caption{Scalograms (spectrograms for wavelets) of the same data using wavelet analysis}
\label{fig:t7}
\end{center}
\end{figure}

In Fig.~\ref{fig:t7}  we show the results for the same data, so one could compare this scalogram with the spectrograms obtained with the former methods. One notices the coherence of the bands during the perturbation.
The wavelet correlations $C'_{\mathrm{v1,v2}}$ are the equivalent of Eq.~\ref{C}, but using the wavelets $W$  instead of the bare signals, In the top two panels of the figure we show $C'_{\mathrm{vi,v1}}$ and $C'_{\mathrm{v2,v2}}$, respectively and in the lower panel we show the wavelet coherence, defined as $(C'_{\mathrm{v1,v2}})^2/(C'_{\mathrm{v1,v1}}C'_{\mathrm{v2,v2}})$. 
Note also the continuous coherence of all bands at all times. 

\begin{figure}[ht!]
\begin{center} 
\includegraphics[width=\columnwidth]{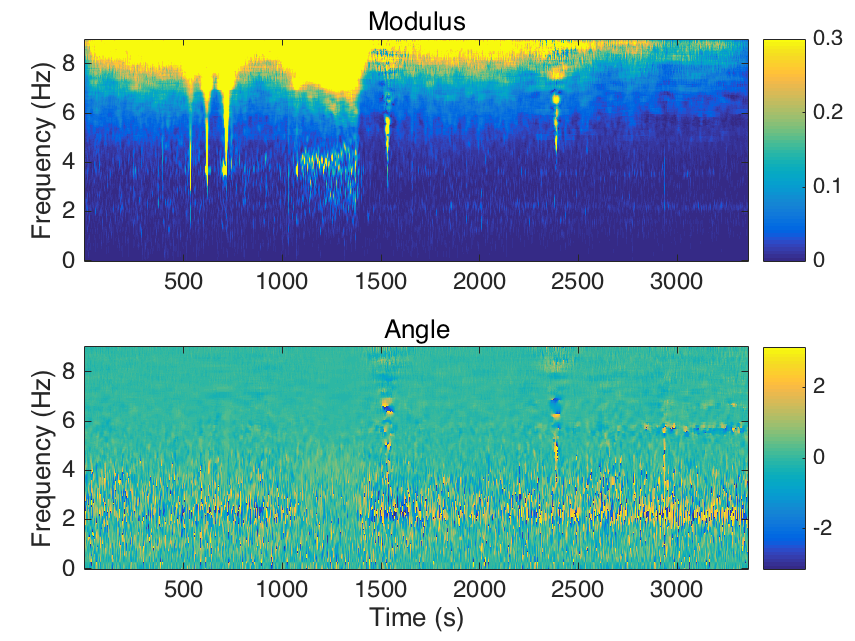}
\caption{Scalograms from the modulus and the phase of  the wavelet cross correlation of the signals. The colour code has been changed to a scale from dark blue to light yellow.}
\label{fig:t8}
\end{center}
\end{figure}

Since the wavelet transforms are complex, the wavelet cross correlation $(C'_{\mathrm{v1,v2}})$ contains information about the modulus and the phase.  In Fig.~\ref{fig:t8} we plot the modulus $\sqrt{[\Re(C'_{\mathrm{v1,v2}})]^2+[\Im(C'_{\mathrm{v1,v2}})^2]}$ on the top and  the phase angle $\arctan[\Re(C'_{\mathrm{v1,v2}})/\Im(C'_{\mathrm{v1,v2}})]$ at the bottom.

Observe that the modulus shows an enhancement of the band at 4 Hz during the perturbation, and that the phase is locked at zero in all bands also in this period. After recovery from the perturbation the phase of the bands at 2 and 6 Hz are locked at values near $\pm \pi$.

In particular, the band at  6 Hz, which is initially not in phase, is gradually locked at $180^{\circ}$  after the challenge \footnote{In unchallenged rats, the theta phase shifts monotonically along the septotemporal axis of the hippocampus \cite{patel}.}. The delay  observing this increase in the phase locking, probably reflects the time necessary for the central vasopressin system to react and reach a new stable state after a hypertonic challenge, in accordance with the results of Ref. \cite{ph}, where it is shown that the vasopressin concentration in plasma after a systemic hyperosmotic challenge increase reaches a maximum  30 minutes after the hypertonic challenge.

\begin{figure}[ht!]
\begin{center} 
\includegraphics[width=\columnwidth]{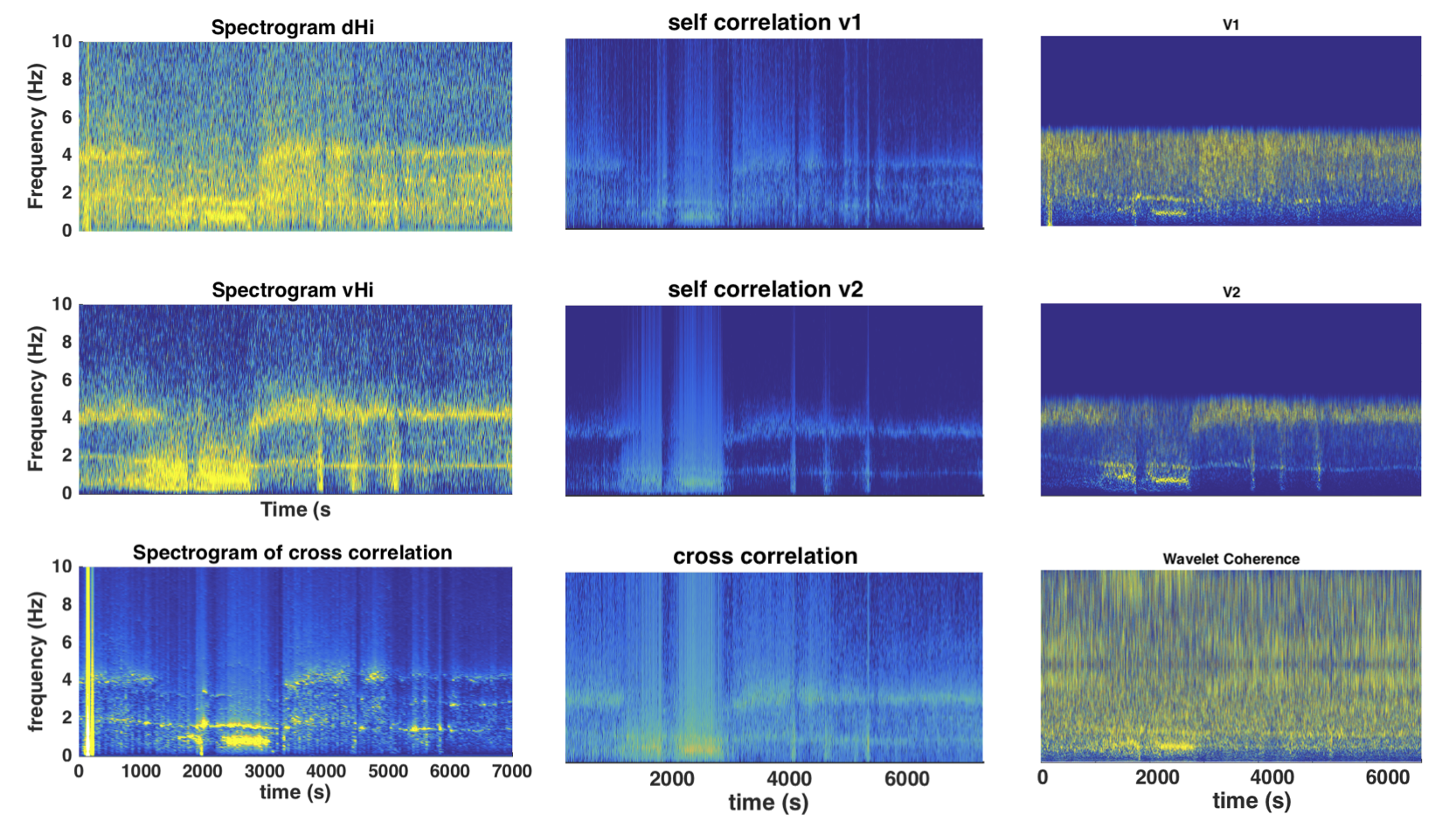}
\caption{Spectrograms based on Fourier (left column), Correlation (central column) and  wavelet analysis (right column). In the left bottom corner we also show the wavelet coherence plot.}
\label{fig:t9}
\end{center}
\end{figure}

In Fig.~\ref{fig:t9} we show a comparison of the methods used for another Data set. The color code has been changed to a scale from dark blue to light yellow for the sake of clarity.

These results suggest that the prolonged activation of the vasopressinergic system can modulate theta activity in the hippocampus and promote the synchronization between the dorsal and ventral regions.
 
\section{THEORETICAL MODEL }
\label{Model}

The experimental results can be understood if one assumes that there are two physiologically  different but  anatomically connected systems in the hippocampus, and that the neurons in these systems are sensitive to the effect of vasopressin.  

One could imagine that the functional connectivity between these two systems is dependent on the localization/nature of the responsive neurons and on the innervation  density/local concentrations of vasopressin in the hippocampus. Therefore, if one produces an external signal that potentiates the release from axon terminals of  osmosensitive neurons located in the hypothalamus , the strength of the connections between the dorsal and ventral hippocampus systems change.
In order to verify this hypothesis, one could build a simple model in which individual neurons are represented as a  set of non linear dynamical equations of the Hodgkin-Huxley type. We propose the following model for an inhibitory neuron \cite{Tiesinga}, which is general enough to serve our purposes,

\begin{equation}
\label{eq1}
C_m\frac{\partial V}{\partial t}=-(I_c+I_L+I_s)\\
\end{equation}
where $C_{m}$ is the capacitance and $V$ is the membrane potential. There are three currents: the calcium ion current,  \[I_{c}=g_{c}mh(V-V_{c}),\]  a leak current   \[I_L=g_L(V-V_L),\]  and a synaptic current\[I_{s}=g_{s}s(V-V_{s}).\] This model could describe a network of neurons connected by synapses of the GABA type. The postsynaptic interaction strength in this model is taken into account by the dynamic variable $s=\sum_is_i/N$, which is the average of all the  $i$ connections between the neuron with all the $N$ members of the network. The kinetic variables obey the following dynamic equations,

 \begin{equation}
\label{eqssh }
\begin{split}
\frac{\partial s }{\partial t}&=k_f F(V)(1-s)-s/t_s\\
\frac{\partial h }{\partial t}&=\frac{(h_{\infty}(V)-h)} {t_{h}(V)}\\
m&=m_\infty (V)
\end{split}
\end{equation}

The  asymptotic values of the kinetic variables are 
\[m_\infty(V)=\frac{1}{1+e^{-(V+40)/7.4}},\]
\[h_\infty(V)=\frac{1}{1+e^{(V+70)/4}},\] the $h$ time scale is
\[t_h(V)/\phi=t_0+t_1/[1+e^{(V+50)/3}],\] where $t_0=30$ and $t_1=500$ are times expressed  in units of $\phi$ which could be adjusted to get  their value in milliseconds.  The presynaptic depolarization  is chosen as
\[F_\infty(V)=\frac{1}{1+e^{-(V+35)/2}},\] so values higher than -35 mV will open the synaptic channels.

The synapsis decay time  $t_s$ must be larger than a certain critical value in order to  have oscillations. The numerical values for the physiological parameters that produce a rhythmic firing pattern are  $C_{m}=1(\mu Fcm^{-2})$,
$g_{L}=0.3$, $g_{c}=3.5$ and $g_{s}=2$, (in $mili\mathrm{Siemens}\;cm^{-2}$),  $t_{s}=16$ (in $ms$), $\phi=0.6$ and $k_{f}=0.5$ (in $ms^{-1}$). The reversal potentials in $mV$ are $V_{L}=-70$, $V_{c}=90$ y $V_{s}=-85$. 

The equations were integrated using a simple Euler method assuring that the time step is small enough for the process to converge. With these parameters a complete period is attained in 500 iterations with a time step of $dt=0.1$. The actual interspike time interval can be adjusted to the experimental data by changing the time scale units of the time step properly. 

\begin{figure} [ht!]
\begin{center}
\includegraphics[width=\columnwidth]{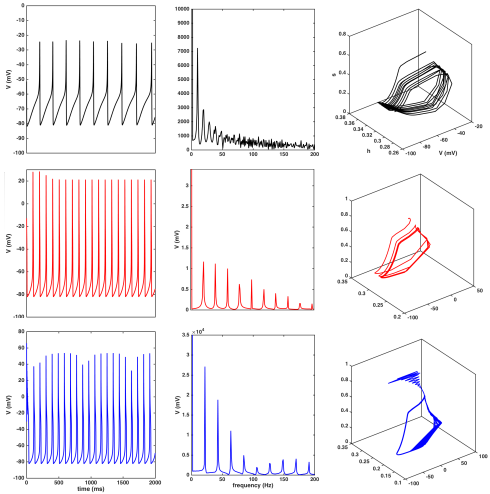}
\end{center}
\caption{From left to right: Membrane Potential of a single neuron as a  function of time, 
Power Spectra and Phase Portrait in the $\{h,V,s\}$ space, for the parameter values in the text and :  (a) $g_{c}=1.23$, (b)$g_{c}=2.50$, (c) $g_{c}=3.50$. }
\label{fig1}
\end{figure}

In Fig.~\ref{fig1} we show the time series of the membrane potential calculated  numerically with this model. Observe that one could change the firing frequency by varying the parameters of the model. In this case we show the variations when using different values of $g_c$. The calculation is for a single neuron with an initial value of the synaptic variable $s_1=0.6392$. In this figure we also show the Fourier spectra and the limit cycles in the ($h,V,s$) space.

Now, we shall use this model to built a network that represents the hippocampus. Given the anatomy of this system, a good model for the oscillations in it would be a linear chain of a number of neuron models connected by synaptic interactions, which should be random, in order to account for alterations in the local field by signals produced in the three dimensional vicinity.  
 
\begin{figure}[ht!]
\begin{center} 
\includegraphics[width=5cm]{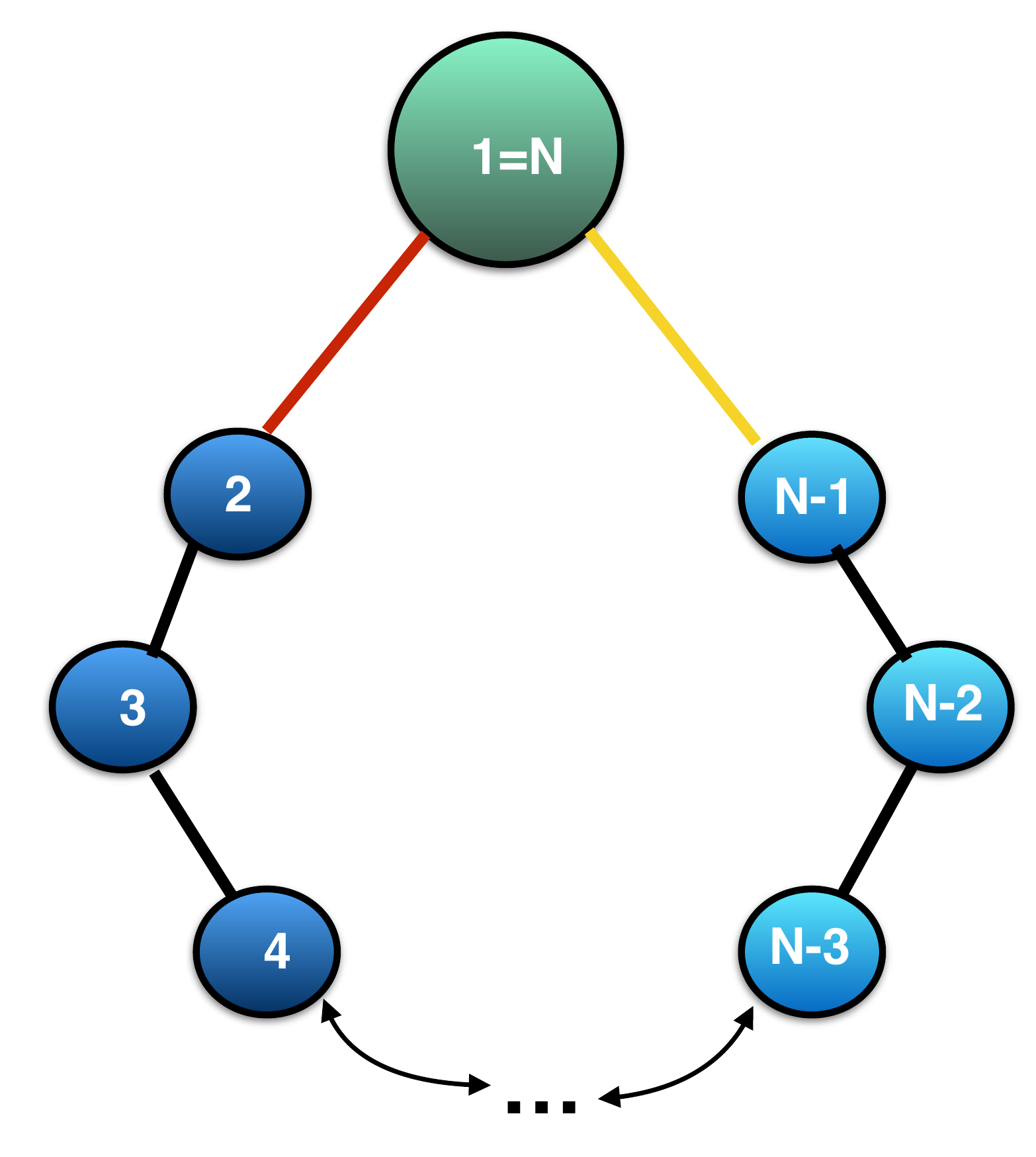}
\caption{Diagram of the model. Each circle represent a measuring location in the hippocampus, all connected by black noisy lines. Neuron 1=N is a signaling neuron in the hypothalamus that is directed to the hippocampus in different locations. The strength of the black, red and yellow connections changes with the perturbation.}
\label{fig2}
\end{center}
\end{figure}

In Fig.~\ref{fig2} we represent these single neuron models as blue circles and their noisy connections with black lines.
Following our basic hypothesis, we need a pace-maker system that fires periodically. This system is represented by the green circle and it is connected to two locations in the chain by interactions of different strength (red and yellow lines in the figure). The strength of these connections may vary when an external event produces a perturbation in the system. This perturbation in the local fields could be measured at different locations simultaneously, in our case, in the experiment one measures two time series of the local fields in different locations of the  CA1/CA2 hippocampal layer. In our model these locations are  the blue circles 2 and N-1.

One now has a system of coupled non linear noisy oscillators, which means that the quantities $V$, $h$, and $s$ are vectors with N entries.

The initial conditions are 
\[V_{inic}=-81.7267;\; s_{inic}=.6392;\;h_{inic}=0.2551,\] for all N systems. 

One defines a specific model by setting the strength of the synaptic interactions between pairs of systems $s_{(i,j)}$. For systems in the chain ($i=2,3,..N-2,N-1$) we define $s(i)= \alpha s_{(i-1,i)}+ \beta s_{(i+1,i)}$. On the cells representing the other regulatory system  one has $s(1)=\alpha_1 s_{(1,1)}+ \beta_1 s_{(2,1)}$ and $s(N)= \alpha_Ns_{(N,N)} + \beta_Ns_{(N-1,N)}$.  The strength of the $\alpha, \beta$  interactions could vary due to the external perturbation at a certain time $t_r-t_p$.

Additionally one should introduce noise in all three dynamical variables to account for the action of the interactions with other cells not considered in the model.
  
\section{Results}

In order to show the sort of results that one obtains with this model, we performed a calculation starting with the synaptic interactions defined above and perturbing the system at certain times. Our goal is to verify if a reasonable hypothesis related with the experimental situation could give similar results.

We assume  that the experimental situation is such that the regions vH and dH are weakly and bidirectionally communicating with each other. A third region in the hypothalamus (PVN)  communicates with the two former regions, with a strength PVN-vH $\gg$ PVN-dh (One detects many more axons  towards vH than to dH). These connections should be modified by the perturbing injection.

We chose a chain with $N=50$ and integrated the dynamical system for  $N_I=2^{17}$ time steps of length $dt=0.5$ ms. These numbers were chosen because the results do not depend crucially on $N$ and they allow to simulate the system for a time lapse comparable to the actual recordings.

We start by adjusting the frequency bands obtained with the model to the ones observed in the unperturbed system. A stability analysis performed in the $(\alpha_i,\beta_i)$ space reveals that the interactions along the chain should be asymmetric, that is, $\alpha \ne \beta$ and that the action of the external cell is also asymmetric. 

Therefore, we start the calculation with the initial conditions $\alpha=0$, $\beta=1.8$, $\alpha_1=1.8$, $\beta_1=0.2$, $\alpha_N=1.8$ and $\beta=0.2$.

We then simulate  the perturbation by  setting $\alpha=1.8$, $\beta=0$, $\alpha_1=1.8$, $\beta_1=0.2$, $\alpha_N=0.2$, and $\beta_N=0.2$ between $t_p=900$ s and $t_r=1300$ s.  Then we restore the system to the original state at time $t_r+dt$. We also introduce white noise of the order of five percent in all  dynamical variables, in order to account for local field variations not taken into account in the chain model.

\begin{figure}[ht!]
\begin{center} 
\includegraphics[width=\columnwidth]{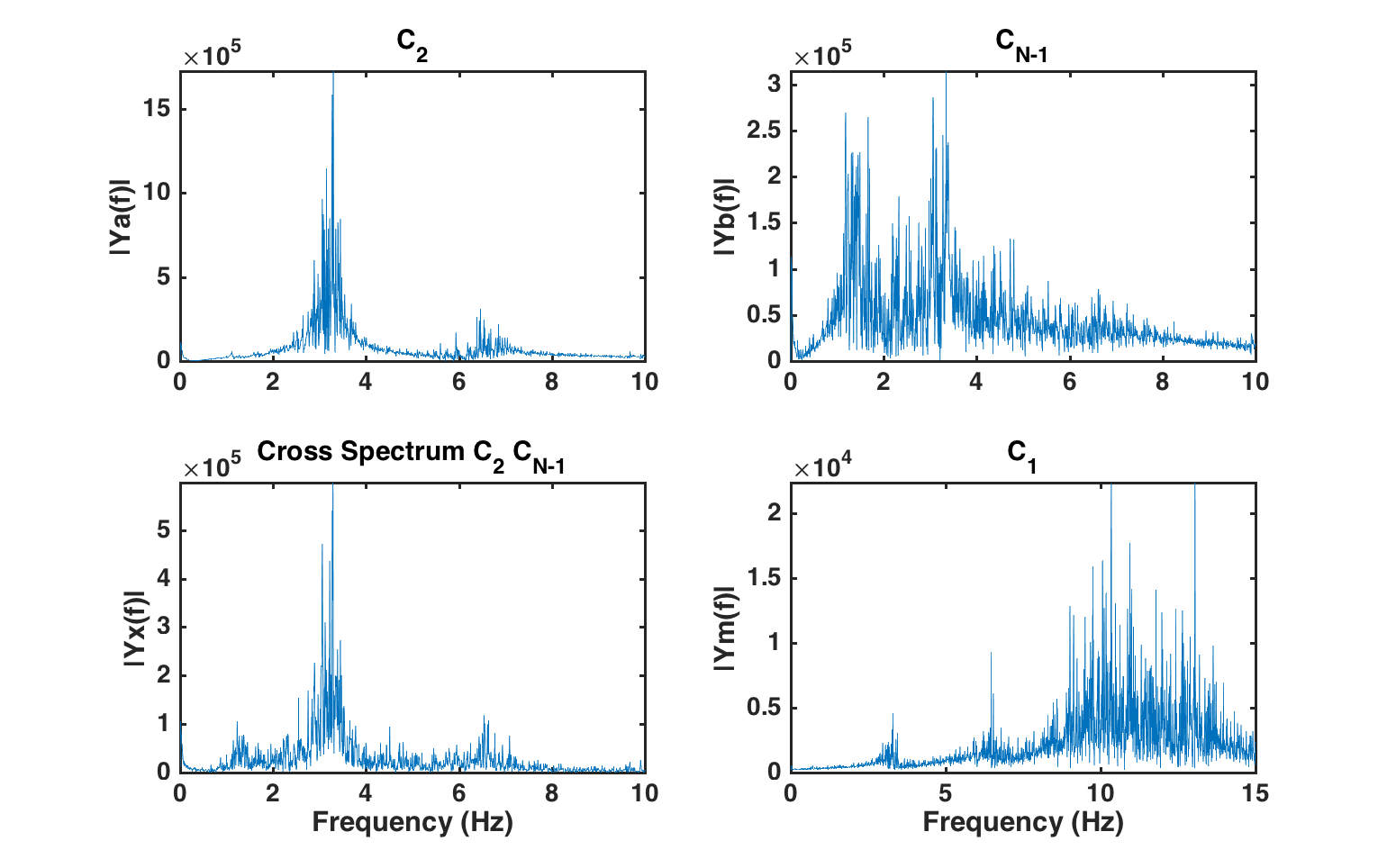}
\caption{ Power spectra of the self correlation of site 2 ($C_2$), site N-1 ($C_{N-1}$), the cross correlation and in the master neuron ($C_1$).}
\label{fig3}
\end{center}
\end{figure}

In Fig.~\ref{fig3} we show the spectra obtained with the correlations. Notice that we catch the main bands appearing in the experimental data shown, with the correct amplitude.  The spectrum of the external  neuron presents a broad band between 10 and 15 Hz.

\begin{figure}[ht!]
\begin{center} 
\includegraphics[width=\columnwidth]{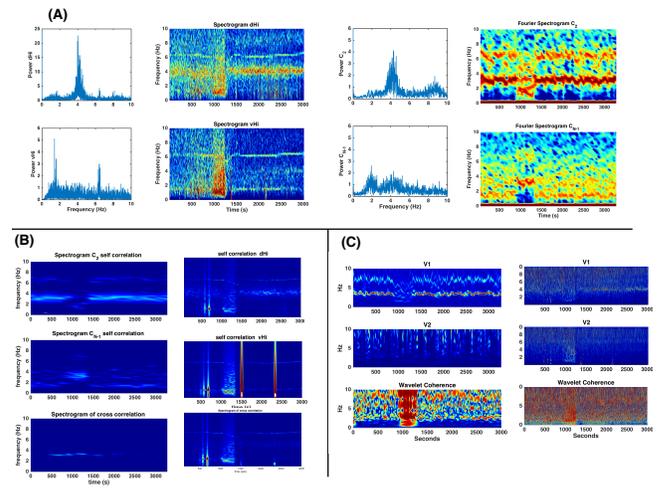}
\caption{ (A) Comparison of the experimental Fourier analysis (on the left)  and from the theoretical model on the right. (B) Same comparison for the correlation analysis and (C) for the wavelet spectrograms.}
\label{fig4}
\end{center}
\end{figure}

The origin of these bands can be investigated by looking at the spectrograms of the analysis of the theoretical time series. In Fig.~\ref{fig4} we show these spectrograms and compare them with the corresponding ones obtained from the experimental Data shown before. 

In panel (A) we show the Fourier analysis, including the spectral bands. Notice the differences with the spectra obtained with correlations. The comparison with the correlation spectrograms is not good because the data contain four external signals not related with the experiment in different times. It is worth noticing  that the wavelet correlation of the delta bands almost disappears during the perturbation and that the wavelet coherence is high everywhere and particularly during the perturbation. The main band at 4 Hz is present only on site v1 ($C_1$) and goes to lower frequency during the perturbation, as in the experiment. The asymmetry  of all bands is also comparable to the experimental data.

 \begin{figure}[ht!]
\begin{center} 
\includegraphics[width=\columnwidth]{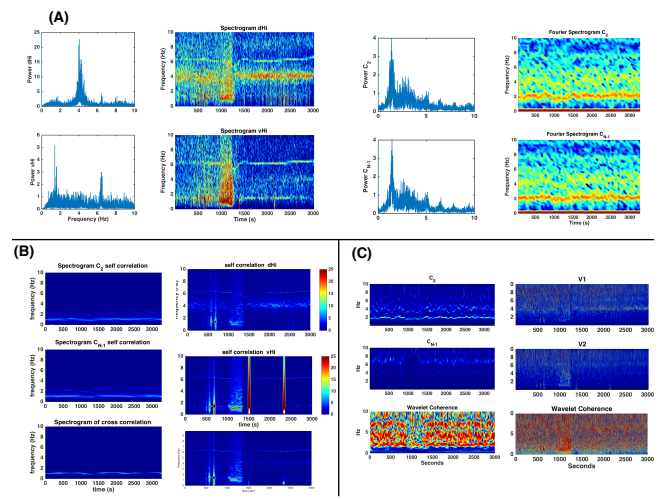}
\caption{ Same comparison of experimental and model results as in Fig. \ref{fig4}}
\label{fig6}
\end{center}
\end{figure}

One could use the results of the model to test if a given hypothesis of the processes that occur in the experiment during a controlled perturbation produces the effects shown in the analysis of the experimental data. 

One could try a different working hypothesis  to interpret the data. For instance, if one proposes that the interactions are symmetric along the chain, that is $s(i)=( s_{i-1,i}+ s_{i+1,i})$, and that the external signals are  quite small $\alpha_1=\alpha_{N-1}=\beta_1=\beta_{N-1}=0.2$. One then assumes that the effect of  the perturbation only consists of increasing the external signal, or $\alpha_1=\alpha_{N-1}=\beta_1=\beta_{N-1}=1$. 

Then one performs the calculations with the same model, but in this case one obtains  very different results, which have very little to do with the experimental findings. This is shown in Fig.~\ref{fig6}. Notice that the model predicts very different results, which means that it could be used to help designing new experiments to validate the hypothesis.

\section{Discussion and Conclusions}

We shall start by discussing the methods used to examine the experimental data.
Fourier power spectra had been successfully used as a first step to detect the main frequencies present in time series data. Fourier transform is a summation over the whole recording time because it assumes stationarity; it does not preserve any information about time. As a consequence, frequencies present for short periods may not appear in the spectrum, or their amplitude may be very small. In the example shown on the left hand side column of  Fig.~\ref{fig:t2} we notice that the principal band in dHi is at 4 Hz while in vHi there are two clear bands at 2 and 6 Hz. These bands are not present all the time. The Fourier windowed transform for creating spectrograms were proposed for analyzing series which change in time. In the spectrograms shown in Fig.~\ref{fig:t2} we find that the band at 2Hz appears in the dHi during the perturbation but it disappears afterwards. the Fourier spectrum for vHi does  show  a $\sim 2$ Hz band which is clearly seen in the spectrogram after the hyperosmotic saline solution injection. 

Calculating the self correlations of the series being examined and analyzing them instead of the row data, will enhance both, Fourier power spectra and the spectrogram, since self correlations will emphasize the real signals. From simultaneous recording it is also possible to investigate if two sites are synchronized or if one influences the other so that the same behavior is seen in the second site but after some delay. This can  be analyzed using the cross correlation of the series, only the synchronous (or delayed) frequencies will appear in the spectrogram. 

Spectrograms can detect relatively long lasting changes. In our example, the perturbation caused by the injection lasted for 400 s, equivalent to 400,000 data, and the spectrograms detect the perturbation very clearly. However, if short lasting changes are present, spectrograms may not detect them. 

Another problem present when windowing a series is that short windows may detect high frequencies, but they may contain little information coming from low frequencies and fail to detect them. In order to solve these problems wavelet analysis has been used. By translating the wavelet using the parameter $b$, time information is kept, while by modifying the scale $a$ different frequencies are detected. Wavelet coherence allows the identification of synchronous signals. In Fig.~\ref{fig:t7} one clearly detects and enhancement of the wavelet correlation between 900 and 1300 s, meaning that the perturbation has caused a synchronization of both signals in the frequency range of interest. 

Using complex wavelets additionally preserves information about the phase at every time measured, in our example we observe a phase locking of all frequencies at zero angle (meaning that signals are synchronized) during the perturbation, confirming the observations of wavelet coherence.

The hippocampus is a major part of the  limbic system that participates in the control of many physiological and behavioral processes. Theta rhythm is the most prominent activity in the hippocampus and is suggested to be involved in the flow of information between hippocampal regions. With the aid of the previous methods for analysis of time series, the effect of an hyperosmotic stimuli that increase the activity of hypothalamic vasopressin neurons, that we have recently shown to innervate the dorsal and the ventral hippocampus can be evaluated, in particular the relationship between theta (4-10 Hz) oscillations of dHi and vHi can be quantified.

The spectrogram of the cross correlation (Fig.~\ref{fig:t6}) shows that at time 1300s (5 minutes after the stimulus, the time necessary to reach a new equilibrium) the band at  6 Hz that initially decreased its frequency  gradually increases until the end of the recording reaching a slightly higher frequency. It is also evident in this figure that after the hypertonic stimulus a  ÒbandÓ around  4 Hz appears, denoting increased correlation between the two time series. Using the more temporally sensitive wavelet coherence analysis, an increase in the coherence of 6 Hz oscillations in the theta range is seen after the hypertonic stimulus, also a decrease in the coherence of delta oscillations ($< $3 Hz) is uncovered. These results suggest that increased activity in the hypothalamus provides a common neuromodulatory input to the hippocampus and enhance the functional coupling in the theta range oscillations between the dorsal and ventral hippocampus.

When sufficient information is available a theoretical model can be proposed based on some reasonable assumptions. Although a model is a simplification of the real phenomena, it can support a deeper understanding about probable ways or mechanisms causing the observed behavior. A model can be used to test different hypothesis, and it can help to choose the most plausible one, or it may also suggest new hypothesis to be experimentally explored

Based in our own anatomical findings regarding  the differential innervation density of vasopressin fibers from the PVN  to the dorsal and ventral hippocampus and  the synaptic innervation of interneurons within the hippocampus \cite{Zhang} as well on the known fact that the dorsal and ventral hippocampus  oscillators differ in frequency, power, and rhythmicity and that the glutamatergic  activation in these nodes can change the direction of information flow along the septotemporal hippocampal axis \cite{gu}, we could make a reasonable hypothesis of the modifications caused by the perturbation. 

A mathematical model  was constructed  to help elucidate  information regarding the interaction dynamics between two neuronal oscillators and a common external input. Within this model a high similitude with the experimental data is obtained if we consider that the dHi and vHi are weakly and bidirectionally connected, that the input from the PVN to the dHi and vHi is also asymmetrical and that when there is a perturbation of the external input, the symmetry of the synaptic interactions between dHi and vHI is reversed. 

Performing a complex behavior relies on the flexible communication between microcircuits located in different brain regions.  A widely accepted assumption is that change in coherence between local theta oscillations can serve as a flexible base for information rerouting.  For instance in our example, the increase in phase locking between dorsal and ventral hippocampus local circuits, might help to coordinate the spatial/cognitive functions associated with the dorsal hippocampus with the emotional/stress associated functions of the ventral hippocampus enhancing cognition in stressful circumstances.  Changes in local oscillatory activity and subsequently in the synchronization level between dorsal and ventral hippocampus can be produced by the synchronic activation of both regions by a common input or by the interaction dynamics within the network connecting both structures. The hypothesis of only perturbing the external signal was tested in our model and the data did not resemble the experimental data, suggesting that this hypothesis could be discarded.

The example discussed and analyzed here clearly illustrates the usefulness of this kind of modeling. Particularly useful are the negative results, or the disagreement between a reasonable hypothesis and the predictions of the model. One can discard immediately the negative ones and use the positive predictions to design new experiments that could  distinguish  between the predictions of two positive models. 

Further studies could take advantage of optogenetics to enhance or diminish with a high temporal and spatial precision the activity of each of the network nodes and reveal mechanisms for synchronous network oscillations.

\section{Acknowledgements} 

LZ and RAB akcnowledge sabbatical grants from DGAPA, UNAM, and are grateful to Prof. Lee Eiden for the hospitality received in the NIMH, USA where this work was completed.

\end{document}